\documentclass[aps,pra,twocolumn,groupedaddress,amsmath,amssymb]{revtex4}
\usepackage{graphicx}
\usepackage{tensor}
\usepackage{xcolor}
\usepackage{mathtools}
\usepackage{mathdots}
\usepackage{bm}

\begin{document}

\title{Diagonalizing the Jaynes-Cummings Hamiltonian and Jaynes-Cummings coherent states}

\author{Stephen M. Barnett$^1$ and Bryan J. Dalton$^2$}

\address{$^1$School of Physics and Astronomy, University of Glasgow, University Avenue, 
Glasgow G12 8QQ, UK}

\address{$^2$Centre for Quantum Science and Technology Theory, School of Science, Computing and Engineering Technology, Swinburne University of Technology, Melbourne, Victoria 3122, Australia}



\begin{abstract} 
We determine the form of the unitary transformation that diagonalizes the Jaynes-Cummings Hamiltonian.
This leads to operators the action of which has a simple interpretation in terms of the dressed states,
the energy eigenstates.  This suggests a set of coherent states and spin coherent states based on the 
dressed states.
\end{abstract}

\maketitle

\section{Introduction}
The Jaynes-Cummings model (universally simplified to the JCM) is sixty years old \cite{JC}.  Its simplicity
and subtle behaviour, as well as its links with experimentally realizable systems 
\cite{HarocheBook,Wineland} have made it one of, perhaps the, most studied models in quantum optics
\cite{BruceBook,Meystre,ShoreKnight,Walls,MTQO,Dutra,Ficek,Agarwal,Milonni}.  Yet in all this work there 
seems not to be the unitary transformation that diagonalizes the Hamiltonian.  In this paper we seek to 
correct this omission.

Finding the dressed states for the JCM is a straightforward exercise because the interaction between the
atom and the field mode preserves the excitation number.  This means that, in the Schr\"odinger 
picture, we only ever couple two states, an excited state with $n$ photons and a ground state with $n+1$
photons.  These two are the only states for which the total number of quanta, $N$, is $n+1$.  The
dressed states are simply superpositions of such coupled bare states.

Addressing the dynamics in the Heisenberg picture is more of a challenge than the simpler 
Schr\"odinger picture treatment, but a complete solution was given by Ackerhalt \cite{Ackerhalt,MTQO}.
His technique was to find superpositions of the photon annihilation and atom lowering operators
that are eigenoperators of the Hamiltonian.  The coefficients in this superposition are also
operators, but are functions only of operator constants of the motion.  We shall find that closely
related ideas to this assist in our derivation of the unitary transformation that diagonalizes the JCM Hamiltonian.

We start with a short discussion of the JCM before, in Section 3, proceeding to derive the form 
of the unitary operator that diagonalizes the JCM Hamiltonian.  In Section 4 we use the
unitary transformation to derive the forms of the dressed operators for the coupled atom and
cavity mode.  Finally, in Section 5, we introduce a new set of coherent states, specifically related
to the JCM and show, in particular, that these tend to the familiar dressed states of the semi-classical model \cite{MTQO} in the appropriate limit.

\section{The JCM: a reminder}
As a brief reminder, but also to set our notation, we give a very brief overview of the JCM
Hamiltonian and of its energy eigenstates.  The JCM Hamiltonian has the form
\begin{equation}
\label{Eq1}
\hat{H} = \frac{\hbar\omega_0}{2}\hat{\sigma}_3 + \hbar\omega\hat{a}^\dagger\hat{a}
+ \hbar\lambda(\hat{a}^\dagger\hat{\sigma}_- + \hat{\sigma}_+\hat{a}) .
\end{equation}
Here $\hat{\sigma}_3 = |e\rangle\langle e| - |g\rangle\langle g|$ and
$\hat{\sigma}_- = |g\rangle\langle e| = \hat{\sigma}_+^\dagger$, where $|e\rangle$ and $|g\rangle$
are, respectively, the excited and ground states, $\hat{a}$ and $\hat{a}^\dagger$ are the usual
photon annihilation and creation operators for the single field mode.  To proceed we transform this into 
an interaction picture and, for simplicity, impose resonance between the atom and the mode so that
$\omega_0 = \omega$.  The resulting interaction picture Hamiltonian is
\begin{equation}
\label{Eq2}
\hat{H}_I = \hbar\lambda(\hat{a}^\dagger\hat{\sigma}_- + \hat{\sigma}_+\hat{a}) .
\end{equation}
The action of this simply exchanges a quantum of energy between the atom and the field.

It is straightforward to determine the energy eigenvalues and eigenstates (dressed states)
for $\hat{H}_I$.  There is a single, zero energy, ground state:
\begin{equation}
\label{Eq3}
\hat{H}_I|g\rangle\otimes|0\rangle  = 0 ,
\end{equation}
where $|0\rangle$ is the vacuum state for the field mode.  The remaining eigenstates are
simple superpositions of states with the same number of quanta:
\begin{equation}
\label{Eq4}
\hat{H}_I|n,\pm\rangle = \pm \hbar\lambda\sqrt{n}|n,\pm\rangle
\end{equation}
where 
\begin{equation}
\label{Eq5}
|n,\pm\rangle = \frac{1}{\sqrt{2}}(|e\rangle\otimes|n-1\rangle \pm |g\rangle\otimes|n\rangle) .
\end{equation}
Here $|n\rangle$ are the photon number eigenstates.  

We note that the JCM has two natural constants of the motion, that is Hermitian operators 
that commute with $\hat{H}_I$.  The first is the interaction Hamiltonian itself and the 
second is 
\begin{equation}
\label{Eq6}
\hat{N} = \hat{\sigma}_+\hat{\sigma}_- + \hat{a}^\dagger\hat{a} ,
\end{equation}
which is the total number of excitations shared between the atom and the field mode. The 
dressed states are simple eigenstates of this quantity:
\begin{align}
\label{Eq7}
\hat{N}|g\rangle\otimes|0\rangle &= 0  \nonumber \\
\hat{N}|n,\pm\rangle &= n|n,\pm\rangle .
\end{align}
We shall find that this second constant of the motion plays a key role in our study.

\section{Unitary transform to diagonalize $\hat{H_I}$}

We start by noting that the generator of the desired unitary transformation cannot be formed from 
a simple product of an annihilation or creation operator and a corresponding lowering or raising
operator for the atom.  It should also commute with the operator $\hat{N}$, corresponding to the 
total excitation number.  To this end we introduce the (anti-Hermitian) operator 
\begin{equation}
\label{Eq8}
\hat{\mathcal{O}} = \hat{\sigma}_+f(\hat{a}^\dagger\hat{a})\hat{a} - 
\hat{a}^\dagger f(\hat{a}^\dagger\hat{a})\hat{\sigma}_- ,
\end{equation}
where $f$ is a real function, to be determined.  The commutation relation between $\hat{\mathcal{O}}$
and $\hat{H}_I$ is
\begin{equation}
\label{Eq9}
[\hat{\mathcal{O}},\hat{H}_I] = 2\hbar\lambda\left(\hat{\sigma}_+\hat{\sigma}_-
\hat{a}\hat{a}^\dagger f(\hat{a}^\dagger\hat{a}) -
\hat{\sigma}_-\hat{\sigma}_+\hat{a}^\dagger\hat{a}f(\hat{a}^\dagger\hat{a}-1)\right) ,
\end{equation}
where we have used the operator identity $f(\hat{a}^\dagger\hat{a})\hat{a} = \hat{a}f(\hat{a}^\dagger\hat{a}-1)$ \cite{Louisell}.  
The fact that the eigenstates of $\hat{H}_I$ depend on the square root of the number of quanta suggests
that we set
$f(\hat{a}^\dagger\hat{a}) = \frac{1}{2}(\hat{a}^\dagger\hat{a} + 1)^{-1/2} = 
\frac{1}{2}(\hat{a}\hat{a}^\dagger)^{-1/2}$ so that 
\begin{equation}
\label{Eq10}
[\hat{\mathcal{O}},\hat{H}_I] =\hbar\lambda\left(\hat{\sigma}_+\hat{\sigma}_-
\sqrt{\hat{a}\hat{a}^\dagger} - \hat{\sigma}_-\hat{\sigma}_+\sqrt{\hat{a}^\dagger\hat{a}}\right) .
\end{equation}
We note that the eigenvalues of this commutator are clearly $\pm\hbar\lambda\sqrt{n}$, as
they must be, and that the eigenstates are the bare, or undressed states:
\begin{align}
\label{Eq11}
[\hat{\mathcal{O}},\hat{H}_I]|g\rangle\otimes|n\rangle &=
-\hbar\lambda\sqrt{n}|g\rangle\otimes|n\rangle \nonumber \\
[\hat{\mathcal{O}},\hat{H}_I]|e\rangle\otimes|n-1\rangle &=
\hbar\lambda\sqrt{n}|e\rangle\otimes|n-1\rangle .
\end{align}
These include the (trivial) ground state:
\begin{equation}
\label{Eq12}
[\hat{\mathcal{O}},\hat{H}_I]|g\rangle\otimes|0\rangle = 0 .
\end{equation}

It remains to construct the unitary transformation based on $\hat{\mathcal{O}}$.  Before addressing this, 
it is interesting to note that we can write $\hat{\mathcal{O}}$ in a somewhat simpler form:
\begin{align}
\label{Eq13}
\hat{\mathcal{O}} &= \hat{\sigma}_+\frac{1}{2}(\hat{a}^\dagger\hat{a} + 1)^{-1/2}\hat{a}
- \hat{a}^\dagger(\hat{a}^\dagger\hat{a} + 1)^{-1/2}\hat{\sigma}_-
\nonumber \\
&= \frac{1}{2}(\hat{\sigma}_+\hat{E} - \hat{E}^\dagger\hat{\sigma}_-) ,
\end{align}
where $\hat{E}$ and $\hat{E}^\dagger$ are the bare lowering and raising operators:
\begin{align}
\label{Eq14}
\hat{E} &= \sum_{n'=0}^\infty(n'+1)^{-1/2}|n'\rangle\langle n'|
\sum_{n=0}^\infty (n+1)^{1/2}|n\rangle\langle n| \nonumber \\
&= \sum_{n=0}^\infty |n\rangle\langle n+1| \nonumber \\
\hat{E}^\dagger &= \sum_{n=0}^\infty |n+1\rangle\langle n| .
\end{align}
These were introduced by Susskind and Glogower \cite{SG} to represent the exponentials of the phase
of the field, although we now know that this is not correct \cite{MTQO,DavidRev,Rodney,PhaseBook}.

To construct the required unitary transformation we note that 
\begin{align}
\label{Eq15}
\hat{\mathcal{O}}^2 &= (-\hat{\sigma}_+\hat{\sigma}_-\hat{E}\hat{E}^\dagger -
\hat{\sigma}_-\hat{\sigma}_+\hat{E}^\dagger\hat{E})/4 \nonumber \\
&= -\frac{1}{4}(\hat{\rm I}\otimes\hat{\rm I} -|g\rangle\langle g|\otimes|0\rangle\langle 0|) ,
\end{align}
which is a quarter of minus the projector onto all the states with the exception of the ground state.
It is now straightforward to find the required unitary operator:
\begin{align}
\label{Eq16}
\hat{U} &= e^{\frac{\pi}{2}\hat{\mathcal{O}}}  \nonumber \\
&= \frac{1}{\sqrt{2}}(\hat{\rm I}\otimes\hat{\rm I} - |g\rangle\langle g|\otimes|0\rangle\langle 0|
+ 2\hat{\mathcal{O}})
+ |g\rangle\langle g|\otimes|0\rangle\langle 0| ,
\end{align}
where the first term couples the excited bare states and the last term leaves the ground state 
unchanged.  We find, by direct calculation, that
\begin{equation}
\label{Eq17}
\hat{U}\hat{H}_I\hat{U}^\dagger = \hbar\lambda(\hat{\sigma}_+\hat{\sigma}_-\sqrt{\hat{a}\hat{a}^\dagger}
-\hat{\sigma}_-\hat{\sigma}_+\sqrt{\hat{a}^\dagger\hat{a}}) .
\end{equation}
We can take advantage of the Pauli operator algebra to write this in a yet more suggestive,
diagonalized, form
\begin{equation}
\label{Eq18}
\hat{H}_{ID} = \hbar\lambda\sqrt{\hat{N}}\hat{\sigma}_3 .
\end{equation}
This is our main result.  Henceforth we denote the transformed (dressed) operators by a 
subscript $D$.

It is clear that the eigenvalues of $\hat{H}_{ID}$ are $\pm\hbar\lambda\sqrt{n}$ and that the
bare atom-field states are its eigenstates:
\begin{align}
\label{Eq19}
\hat{H}_{ID}|g\rangle\otimes|n\rangle &= -\hbar\lambda\sqrt{n}|g\rangle\otimes|n\rangle
\nonumber \\
\hat{H}_{ID}|e\rangle\otimes|n-1\rangle &= \hbar\lambda\sqrt{n}|e\rangle\otimes|n-1\rangle .
\end{align}
We can also write the operator $\hat{\mathcal{O}}$ in a simpler form:
\begin{equation}
\label{Eq20}
\hat{\mathcal{O}} = \frac{1}{2\sqrt{\hat{N}}}(\hat{\sigma}_+\hat{a} - \hat{a}^\dagger\hat{\sigma}_-) ,
\end{equation}
where we have used the fact that $\hat{N}$ commutes with $\hat{\sigma}_+\hat{a}$ and with $\hat{a}^\dagger\hat{\sigma}_-$.

\section{Dressed operators}
To complete our analysis we should determine the forms of the atom and field operators in the 
transformed picture and provide their physical significance.  The required dressed annihilation 
and lowering operators are
\begin{align}
\label{Eq21}
\hat{a}_D &= \hat{U}\hat{a}\hat{U}^\dagger \nonumber \\
\hat{\sigma}_{-D} &= \hat{U}\hat{\sigma}_-\hat{U}^\dagger .
\end{align}
We could determine the forms of these in terms of the original undressed operators, but it is 
both simpler and more informative to consider their action on the dressed states.

We start with the photon annihilation operator, the action of which on the bare states is
\begin{align}
\label{Eq22}
\hat{a}|g\rangle\otimes|n\rangle &= \sqrt{n}|g\rangle\otimes|n-1\rangle \nonumber \\
\hat{a}|e\rangle\otimes|n\rangle &= \sqrt{n}|e\rangle\otimes|n-1\rangle 
\end{align}
for $n \geq 1$.  It follows that 
\begin{align}
\label{Eq23}
\hat{a}_D\hat{U}|g\rangle\otimes|n\rangle &= \hat{U}\hat{a}\hat{U}^\dagger\hat{U}|g\rangle\otimes|n\rangle
= \sqrt{n}\hat{U}|g\rangle\otimes|n-1\rangle \nonumber \\
\hat{a}_D\hat{U}|e\rangle\otimes|n\rangle &= \sqrt{n}\hat{U}|e\rangle\otimes|n-1\rangle ,
\end{align}
with $\hat{a}_D\hat{U}|e\rangle\otimes|0\rangle = 0 = \hat{a}_D\hat{U}|g\rangle\otimes|0\rangle$.
Hence we require only the forms of the transformed states $\hat{U}|g\rangle\otimes|n\rangle$ and
$\hat{U}|e\rangle\otimes|n\rangle$:
\begin{align}
\label{Eq24}
\hat{U}|g\rangle\otimes|0\rangle &= |g\rangle\otimes|0\rangle \nonumber \\
\hat{U}|g\rangle\otimes|n\rangle &= |n,+\rangle \nonumber \\
\hat{U}|e\rangle\otimes|n\rangle &= |n+1,-\rangle ,
\end{align}
for $n \geq 1$.  We can simplify these by \emph{defining} the state 
\begin{equation}
\label{Eq25}
|0,+\rangle \equiv |g\rangle\otimes|0\rangle ,
\end{equation}
so that the second line of Eq. (\ref{Eq24}) applies for all excitation numbers $n$.  Note that there 
is no $|0,-\rangle$ state.

It now follows that the dressed annihilation operator acts, simply, to reduce the total number 
of excitations in both the symmetric and the antisymmetric dressed states:
\begin{align}
\label{Eq26}
\hat{a}_D|n,+\rangle &= \sqrt{n}|n-1,+\rangle  \nonumber \\
\hat{a}_D|n,-\rangle &= \sqrt{n-1}|n-1,-\rangle .
\end{align}
Hence the dressed annihilation operator is
\begin{equation}
\label{Eq27}
\hat{a}_D = \sum_{n=1}^\infty\sqrt{n}|n-1,+\rangle\langle n,+| + 
\sum_{n=1}^\infty \sqrt{n-1}|n-1,-\rangle\langle n,-| .
\end{equation}
It is straightforward to confirm that the commutation relation with the corresponding
dressed creation operator has the form
\begin{equation}
\label{Eq28}
[\hat{a}_D,\hat{a}_D^\dagger] = \hat{\rm I}\otimes\hat{\rm I} ,
\end{equation}
as must be the case as $\hat{a}_D$ and $\hat{a}_D^\dagger$ are obtained from 
$\hat{a}$ and $\hat{a}^\dagger$ via a unitary transformation.

Let us turn now to the atom lowering operator for which we have the actions
\begin{align}
\label{Eq29}
\hat{\sigma}_-|g\rangle\otimes|n\rangle &= 0 \nonumber \\
\hat{\sigma}_-|e\rangle\otimes|n\rangle &= |g\rangle\otimes|n\rangle .
\end{align}
It then follows that the dressed operator has the actions 
\begin{align}
\label{Eq30}
\hat{\sigma}_{-D}\hat{U}|g\rangle\otimes|n\rangle &= 
\hat{U}\hat{\sigma}_-\hat{U}^\dagger\hat{U}|g\rangle\otimes|n\rangle = 0 \nonumber \\
\hat{\sigma}_{-D}\hat{U}|e\rangle\otimes|n\rangle &= \hat{U}|g\rangle\otimes|n\rangle
\end{align}
for all $n$ and hence we can write $\hat{\sigma}_{-D}$ in the form
\begin{equation}
\label{Eq31}
\hat{\sigma}_{-D} = \sum_{n=0}^\infty |n,+\rangle\langle n+1,-| ,
\end{equation}
which clearly satisfies the requirement that $\hat{\sigma}_{-D}^2 = 0$.  It is simple to 
confirm that the dressed lowering operator and its conjugate satisfy the required anticommutation
relation:
\begin{equation}
\label{Eq32}
\{\hat{\sigma}_{-D},\hat{\sigma}_{+D}\} = \hat{\rm I}\otimes\hat{\rm I} .
\end{equation}
We note also that $\hat{N}$ is a constant of the motion and commutes with $\hat{\mathcal{O}}$
and it follows that this quantity is invariant under the unitary transformation:
\begin{equation}
\label{Eq33}
\hat{N}_D = \hat{\sigma}_{+D}\hat{\sigma}_{-D} + \hat{a}_D^\dagger\hat{a}_D = \hat{N} .
\end{equation}

The two dressed lowering operators, one for the field and one for the atom, have simple effects on the 
dressed states.  The dressed annihilation operator, $\hat{a}_D$, reduces by one the number of quanta
in a dressed state while retaining the symmetry of the state; a symmetric state $(+)$ is transformed into
a symmetric state and an antisymmetric one $(-)$ into an antisymmetric one.  The dressed atom lowering 
operator also reduces the total number of excitations by one but also changes an antisymmetric state
into a symmetric one.  Acting on a symmetric state it gives zero.  This situation is reminiscent of
Mollow's treatement of resonance fluorescence \cite{Mollow}, see also \cite{CT}, in which a triplet of fluorescent 
spectral lines can be observed given sufficiently strong driving.  The central line, at the natural
transition frequency may be compared with the action of $\hat{a}_D$, while the observed lower
frequency $(\omega - \lambda)$ may be associated with the action of $\hat{\sigma}_{-D}$.  The 
remaining line (centered at frequency $\omega+\lambda$) may be associated with the more complicated
action of $\hat{a}_D^2\hat{\sigma}_{+D}$.

\section{JCM coherent states}
The coherent states were introduced by Glauber in connection with the coherence properties of 
light \cite{Glauber,GlauberBook} and have since played a crucial role in contrasting quantum effects
from those that have a (semi-)classical explanation \cite{MandelWolf}.  Although intrinsically
associated with electromagnetic field modes, with the associated harmonic oscillator algebra, it
was quickly realized that similar states could be formed for other quantum systems 
\cite{Perelomov,Klauder}.  In this section we consider the forms of coherent states associated with
the properties of our dressed operators, $\hat{a}_D$ and $\hat{\sigma}_{-D}$.  We find that there are
clear similarities between these and those more commonly associated with a single field mode or a 
single atom.  We should note that the JCM has been studied using a combination of regular coherent states 
for the field and Grassmann coherent states for the atom \cite{BryanJCM}.  The coherent states introduced 
in this section are distinct from these, but both sets are aimed at introducing a useful description of
the properties and dynamics of the JCM.

We start with the coherent states associated with the annihilation operator.  Recall that for a 
single (bare) oscillator, we can define a coherent state by means of a unitary operator acting
on the ground or vacuum state:
\begin{equation}
\label{Eq34}
|\alpha\rangle = \exp(\alpha\hat{a}^\dagger - \alpha^*\hat{a})|0\rangle = 
e^{-|\alpha|^2/2}\sum_{n=0}^\infty \frac{\alpha^n}{\sqrt{n!}}|n\rangle .
\end{equation}
The quasi-classical behaviour of the mode prepared in this state may be traced to the fact that the 
coherent states are right-eigenstates of the annihilation operator:
\begin{equation}
\label{Eq35}
\hat{a}|\alpha\rangle = \alpha|\alpha\rangle .
\end{equation}
The operator $\hat{a}_D$ acts on both the atom and the field mode and there are two effective
ground states.  The coherent states are
\begin{align}
\label{Eq36}
|\alpha,+\rangle & = \exp(\alpha\hat{a}^\dagger - \alpha^*\hat{a})|0,+\rangle
= e^{-|\alpha|^2/2}\sum_{n=0}^\infty\frac{\alpha^n}{\sqrt{n!}}|n,+\rangle \nonumber \\
|\alpha,-\rangle & = \exp(\alpha\hat{a}^\dagger - \alpha^*\hat{a})|0,-\rangle
= e^{-|\alpha|^2/2}\sum_{n=0}^\infty\frac{\alpha^n}{\sqrt{n!}}|n+1,-\rangle .
\end{align}
Note that the state $|\alpha,+\rangle$ is similar to that for the familiar single-mode 
coherent state, Eq. (\ref{Eq34}), but that the state $|\alpha,-\rangle$ is a superposition
of states starting with $|1,-\rangle$, which is the lowest energy antisymmetric state.
These two coherent states reflect, naturally, the existence of two distinct coherent states,
$|g\rangle\otimes|\alpha\rangle$ and $|e\rangle\otimes|\alpha\rangle$, for the bare atom
and field mode.  As with the bare coherent states, both $|\alpha,+\rangle$ and $|\alpha,-\rangle$
are right eigenstates of $\hat{a}_D$:
\begin{align}
\label{Eq37}
\hat{a}_D|\alpha,+\rangle &= \alpha|\alpha,+\rangle \nonumber \\
\hat{a}_D|\alpha,-\rangle &= \alpha|\alpha,-\rangle .
\end{align}
The two sets of states are mutually orthogonal,
\begin{equation}
\label{Eq37a}
\langle\alpha,+|\alpha',-\rangle = 0 ,
\end{equation}
and together they form an overcomplete representation of the atom-field state space:
\begin{equation}
\label{Eq37b}
\int \frac{d^2\alpha}{\pi} |\alpha,+\rangle\langle\alpha,+| + |\alpha,-\rangle\langle\alpha,-|
= \hat{\rm I}\otimes\hat{\rm I} .
\end{equation}
Both of the JCM coherent states are entangled states of the atom and of the field.  As the magnitude of
$\alpha$ increases, however, the degree of entanglement reduces and tends, in the limit, to analogues
of the semiclassical dressed states, with the atom driven by a classical (c-number) field:
$2^{-1/2}(|e\rangle \pm |g\rangle)|\alpha\rangle$.

The spin, or atomic, coherent states \cite{MTQO,Radcliffe,Arrechi}, like their oscillator counterparts,
can be parameterized by a single complex number.  For our two-level atom, we can write this in the form
\begin{equation}
\label{Eq38}
|\zeta\rangle = (1+|\zeta|^2)^{-1/2}(|g\rangle + \zeta|e\rangle) .
\end{equation}
We note that all pure states of the atom can be written in this form.  The spin coherent state is 
not an eigenstate of the lowering operator, $\hat{\sigma}_-$, but it does have the non-zero
expectation value
\begin{equation}
\label{Eq39}
\langle\zeta|\hat{\sigma}_-|\zeta\rangle = \frac{\zeta}{1+|\zeta|^2}.
\end{equation}
It is straightforward to obtain the analogous coherent states for the dressed operators $\hat{\sigma}_{-D}$.
As with the coherent states for $\hat{a}_D$, there will be more than one of these; there are, in fact,
infinitely many with one associated with each pair of dressed states $|n,+\rangle$ and $|n+1,-\rangle$.
For these two energy eigenstates we have the spin coherent state
\begin{equation}
\label{Eq40}
|\zeta,n\rangle = (1+|\zeta|^2)^{-1/2}(|n,+\rangle + \zeta|n+1,-\rangle) .
\end{equation}
This is not an eigenstate of $\hat{\sigma}_{-D}$ but has the expectation value
\begin{equation}
\label{Eq41}
\langle\zeta,n|\hat{\sigma}_-|\zeta,n\rangle = \frac{\zeta}{1+|\zeta|^2}.
\end{equation}
The oscillator and spin coherent states are a natural consequence of the unitary connection between 
our bare and dressed states and operators.  In particular we have
\begin{align}
\label{Eq42}
|\alpha,+\rangle &= \hat{U}|g\rangle\otimes|\alpha\rangle \nonumber \\
|\alpha,-\rangle &= \hat{U}|e\rangle\otimes|\alpha\rangle \nonumber \\
|\zeta, n\rangle &= \hat{U}|\zeta\rangle\otimes|n\rangle .
\end{align}
This means that the well-known properties and applications of the coherent states can readily 
be extended to the JCM dressed states.

\section{Conclusion}
Part of the appeal of the JCM, certainly to theorists, is the comparative simplicity with which it 
can be analyzed, coupled with the richness of the associated dynamics.  There is scarcely a feature
of quantum optics that has not been explored using the JCM, with important examples including 
collapses and revivals \cite{Eberly}, the generation of squeezed light \cite{Wodkiewicz,PhysScrip}
and the generation of Schr\"{o}dinger cat states \cite{GeaB}.

We have derived the unitary transformation that diagonalizes the Jaynes-Cummings Hamiltonian.  The
full Hamiltonian in its diagonalized form is
\begin{equation}
\label{Eq43}
\hat{H}_D = \hbar\omega\hat{N} + \hbar\lambda\sqrt{\hat{N}}\hat{\sigma}_3 .
\end{equation}
It is immediately clear that the spectrum of the energy eigenvalues is $\hbar\omega n \pm \hbar\lambda\sqrt{n}$, which corresponds to the single zero-energy ground state and then pairs of 
near-degenerate states split by the interaction term.

Having the unitary transformation means that we can transform the atom and field operators into
operators that act directly on the basis of the dressed states.  We have constructed the annihilation
and atom lowering operators for the dressed Hamiltonian, $\hat{a}_D$ and $\hat{\sigma}_{-D}$,
and explored some of their properties.  It is to be hoped that these may find further application
in the next sixty years of the JCM.

\begin{acknowledgments}

We am grateful to the Royal Society for the award of a Research Professorship.  We are grateful
to Sarah Croke and James Cresser for helpful comments.  This work was supported by the Royal 
Society, grant number RP150122

\end{acknowledgments}


\end{document}